\documentstyle[twocolumn,prl,aps,graphicx]{revtex}
\begin{document}

\draft
% for two column  activate the line below...                
\twocolumn[\hsize\textwidth\columnwidth\hsize\csname@twocolumnfalse\endcsname

\title{Double sign reversal of the vortex Hall effect in $\rm{\bf{YBa_2Cu_3O_{7-\delta}}}$
thin films in the strong pinning limit of low magnetic fields}

\author{W. G\"ob$^1$, W. Liebich$^1$, W. Lang$^{1,4,*}$, I. Puica$^2$,
Roman Sobolewski$^{3,\dagger}$, R. R\"ossler$^4$, J.~D.~ Pedarnig$^4$,
D. B\"auerle$^4$ }
\address{$^1$Institut f\"ur Materialphysik der Universit\"at Wien,
Kopernikusgasse 15, A-1060 Wien, Austria \\
$^2$Physics Department, Polytechnic University of Bucharest,
77206 Bucharest 6, Romania \\
$^3$Department of Electrical and Computer Engineering and Laboratory for
Laser Energetics, University of Rochester, Rochester, NY 14627-0231, USA \\
$^4$Angewandte Physik, Johannes-Kepler-Universit\"at Linz,
A-4040 Linz, Austria}

\date{\today}
\maketitle

\begin{abstract} Measurements of the Hall effect and the resistivity in twinned
YBa$_2$Cu$_3$O$_{7-\delta}$ thin films in magnetic fields $B$ oriented
parallel to the crystallographic $c$-axis and to the twin boundaries reveal a
double sign reversal of the Hall coefficient for $B \le 1$\ T. In high transport
current densities, or with $B$ tilted off the twin boundaries by $5^\circ$,
the second sign reversal vanishes.
The power-law scaling of the Hall conductivity to the longitudinal
conductivity in the mixed state is strongly modified in the regime of the
second sign reversal. Our observations are interpreted as strong,
disorder-type dependent vortex pinning and confirm that the Hall
conductivity in high temperature superconductors is not independent
of pinning.

\end{abstract}

% for two column  activate the line below...                            
]                

The unusual behavior of the Hall effect in many high-temperature and in
some conventional superconductors in the mixed state attracts
considerable interest.
In particular, the sign reversal of the Hall angle below 
the critical temperature $T_c$, as compared to the normal state,
is in contrast to traditional models for the vortex Hall effect and is
regarded as a fundamental problem of vortex dynamics. 
Several theoretical approaches have attempted to explain this
phenomenon, but no agreement has been achieved.
The questions, whether the Hall anomaly is an intrinsic
electronic property, determined by the trajectory of an individual vortex
\cite{FERR92,VANO95,KHOM95,KOPN96} if collective vortex phenomena
are essential \cite{AO98,JENS92}, or if vortex pinning is indispensable
for the sign reversal \cite{WANG91,WANG94}, are currently not resolved.
Other models are based on the general grounds of the
time-dependent Ginzburg-Landau theory
 \cite{DORS92,TROY93,KOPN93,NISH97}, but one needs a microscopic
theory to predict the sign of the vortex Hall effect. 

The experiments revealed that the Hall anomaly in high temperature
superconductors (HTS), that is only observed in moderate magnetic fields,
becomes more prominent in smaller magnetic fields, and attains its maximum
within the vortex liquid and thermodynamic fluctuation range
\cite{CHIE91,LIU97}. Above $T_c$, a rapid drop of the Hall resisitivity $\rho_{yx}$
preceeds its sign reversal \cite{LANG94}. The occurence of the Hall anomaly
appears to be connected with the carrier concentration  \cite{NAGA98},
being absent in heavily overdoped cuprates.
In highly anisotropic HTS, like Bi$_2$Sr$_2$CaCu$_2$O$_x$, a double sign
reversal is observed \cite{ZAVA91} that is attributed to weak pinning
in these cuprates. This latter observation raised the question, whether such
double sign reversal could also exist in YBa$_2$Cu$_3$O$_{7-\delta}$ (YBCO). 
In fact, a positive Hall effect has been observed in YBCO far below $T_c$
when pinning is overpowered by intense measurement tranport
currents \cite{NAKA98}.

Another issue is the scaling of the transverse (Hall) resistivity to the
longitudinal resistivity $|\rho_{yx}| \propto \rho_{xx}^\beta$ that can be
experimentally observed with $\beta \sim 1.7$ \cite{LUO92}, irrespective
of the sign of  $\rho_{yx}$ \cite{SAMO93}. Theoretically, an universal scaling
law was derived near a vortex-glass transition \cite{DORS92a},
or, as a general feature of the disorder-dominated vortex dynamics,
with the specific prediction of $\beta = 2$  \cite{VINO93,LIU95}.
The latter model also concluded that the Hall conductivity
$\sigma_{xy}=\rho_{yx}/(\rho_{xx}^2+\rho_{yx}^2)$ should be independent
of pinning, in sharp contrast with theories that originate the Hall anomaly
on pinning \cite{WANG91,WANG94}. Several groups have reported experiments
on samples with artificially introduced defects, and their data have been
interpreted both in favor and against the pinning independence of
$\sigma_{xy}$ \cite{SAMO95,KANG96}. 
Measurements on pure YBCO single crystals revealed a sharp change of
$\sigma_{xy}$ when crossing through the melting transition \cite{DANN98}.
Finally, recent theoretical results have suggested that $\sigma_{xy}$ can be
influenced by strong pinning, eventually leading to the Hall effect with
opposite sign, as compared to its value in the flux-flow regime \cite{KOPN99},
and that the presence or absence of such effect depends on the
dimensionality of the pinning centers near Bose or vortex glass transition,
respectively \cite{IKED99}.
 
In this paper, we report measurements of the vortex Hall effect in
YBCO thin films in a large range of magnetic fields, with particular
emphasis put on small magnetic fields. Our studies allow us to investigate
the Hall effect in the limit of strong pinning on twin boundaries with a
dilute vortex density, as opposed to previous experiments, where pinning
was enhanced by radiation damage. The latter may cause spurious effects,
such as amorphous regions and local changes of the oxygen content along
the heavy-ion tracks. The low-field results are augmented by pulsed high
transport current density and oblique magnetic field measurements of
the vortex Hall effect.    

We present data collected from 100-nm-thick, epitaxial YBCO films, 
deposited either by single-target rf sputtering on ${\rm LaAlO_3}$ substrates
or by pulsed-laser deposition on MgO substrates. In both cases, the films were
highly epitaxial with the onset of superconducting transition at 90 K and with
critical current densities $j_c > 3$~MA/cm$^2$ at 77~K. The experiments were
performed with 17-Hz ac currents at $j=250$\ A/cm$^2$.
The measurements from 1 to 13 T were made in a superconducting solenoid
using a standard cryogenic technique, while the low-field measurements
from 32 mT to 1 T were performed in a closed-cycle cryocooler and with an
electromagnet. The high sensitivity was achieved by fabricating thin film
structures with the excellent alignment of Hall probes.  Particular care was
exercised to exclude spurious signals from the earth's magnetic field and
the remanence of the magnet's pole pieces. To overcome the pinning,
measurements with 3-$\mu$s-long, high-current density pulses were performed
using a four-probe method, fast differential amplifier, and voltage detection
by a boxcar averager. The temperature rise of the sample relative to the bath
was smaller than 1 K and was always corrected using the YBCO film as
an intrinsic thermometer \cite{KUNC94}. 

\begin{figure}[!tb]
\includegraphics[clip,width=6.5cm]{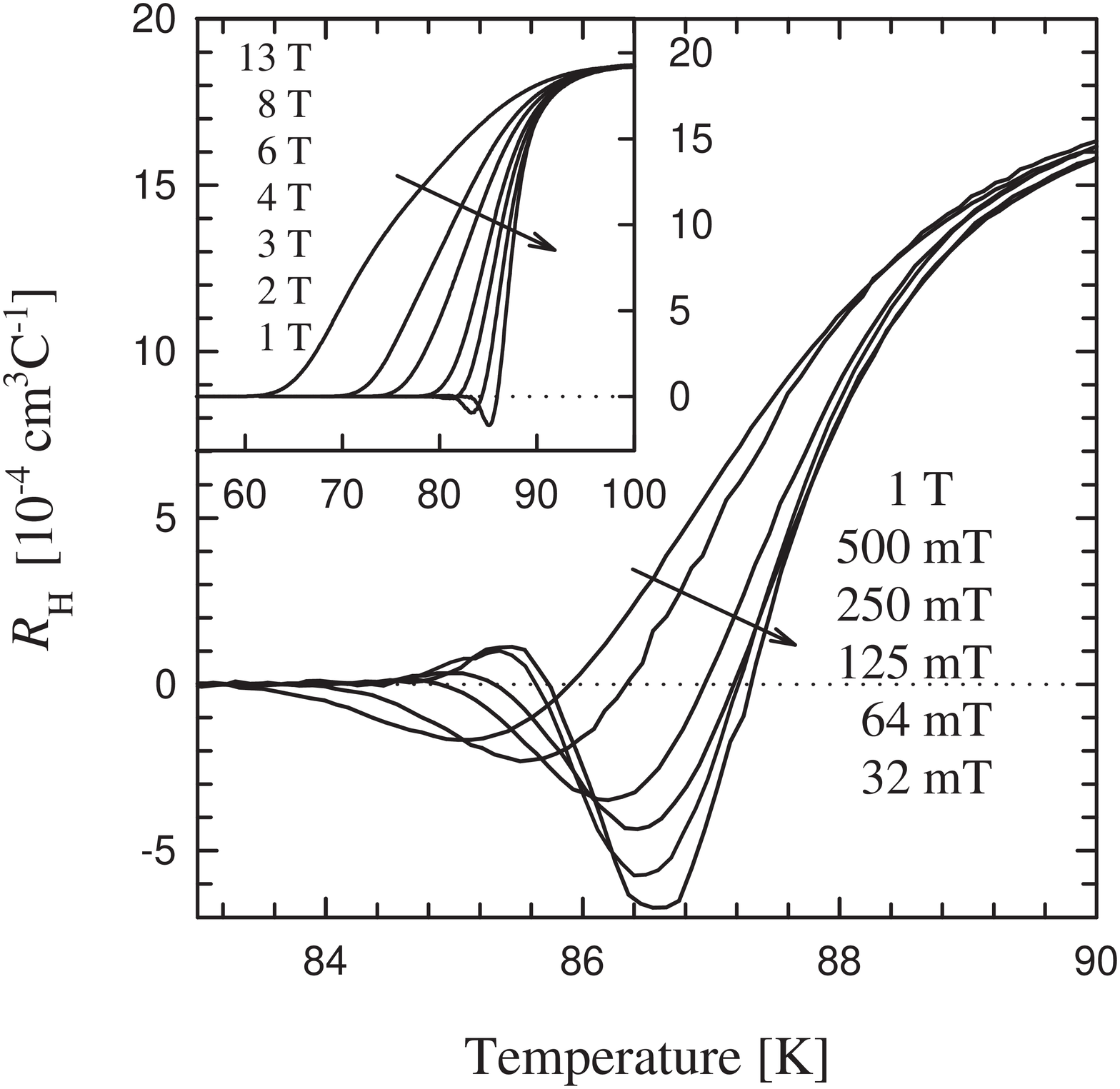}
%\vspace{1cm} 
\caption{Temperature dependence of the Hall coefficient of a YBCO thin film
for various magnetic fields $B$ parallel to the film $c$-axis. The inset shows
the conventional high-$B$ field dependence of $R_H$ on temperature.}
\label{fig:Hall}
\end{figure}

Resistivity of the YBCO thin film deep in the mixed state with $B$ parallel to
$c$-axis is smaller than that in single crystals due to enhanced pinning at
defects and does not show a first-order transition at low temperatures.
At the same time, the shape of the upper part of the transition curve is common
to both thin films and single crystals \cite{SART97} and can be well
characterized by renormalized superconducting-order-parameter
fluctuations \cite{IKED91}.
Figure~\ref{fig:Hall} displays the Hall coefficient $R_H$ of a YBCO film for a
wide range of magnetic fields from 32 mT to 13 T. It demonstrates that $R_H$
is positive at all temperatures for $B\geq 4$\ T (inset) and reverses its sign at
lower fields. The negative sign Hall anomaly increases significantly when the
magnetic field is reduced below 1~T.
Simultaneously, the very surprising new result, observed as a second sign
reversal of $R_H$, can be clearly identified below 0.25 T. This finding
contradicts the notion \cite{NAKA98} that the double sign reversal can only
be observed in weak pinning. On the contrary, we will show that the double
sign reversal of $R_H$ presented in Fig.\ \ref{fig:Hall} is a result of strong
pinning of vortices. 

\begin{figure}[tb]
\includegraphics[clip,width=6.5cm]{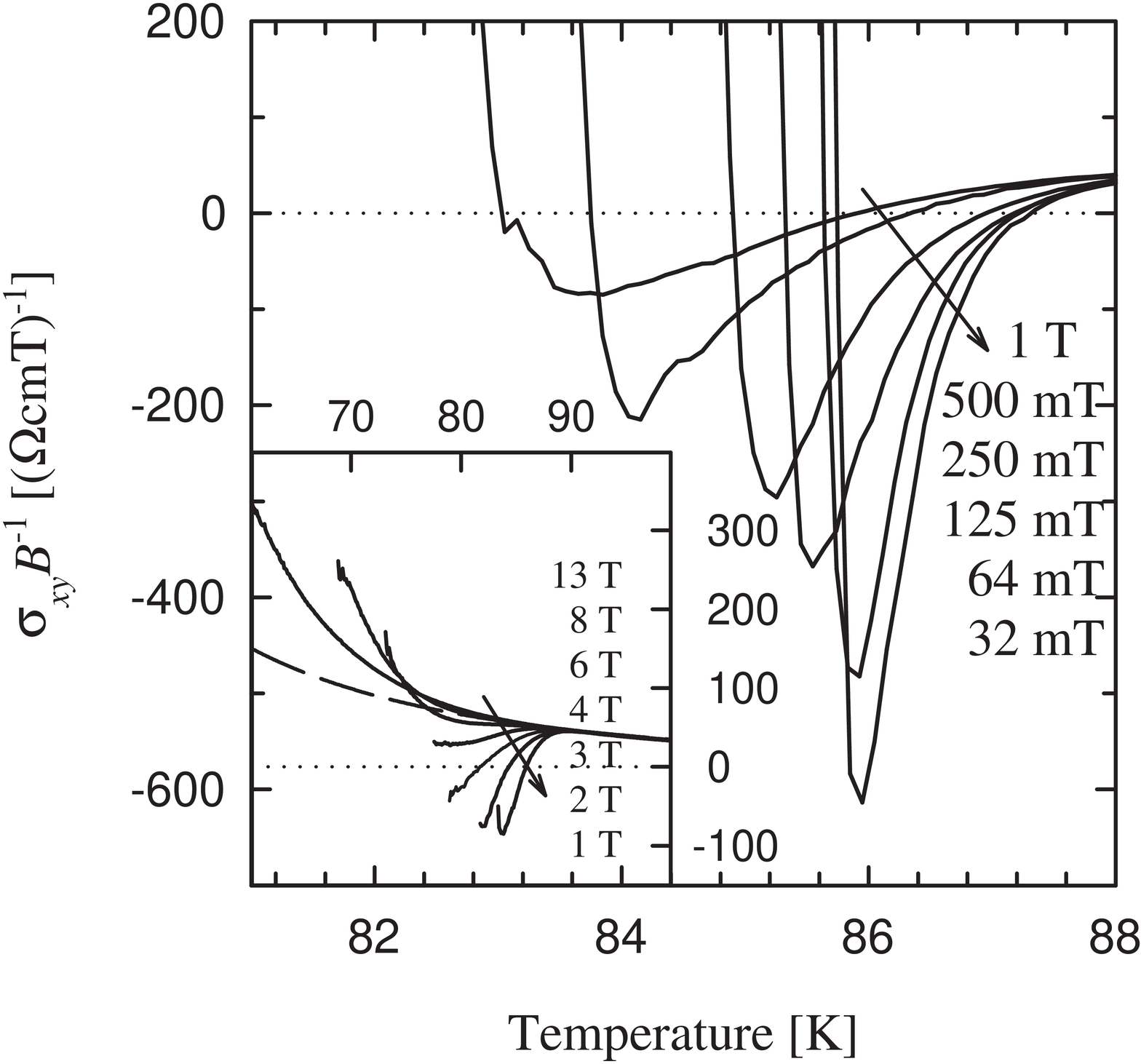}
\caption{$B$-field-normalized Hall conductivity of YBCO dependence on
temperature for various magnetic fields $B$ parallel to the film $c$-axis.
The inset shows the conventional high-$B$ field dependence of
$\sigma_{xy}B^{-1}$ on temperature, and the broken line represents an
extrapolation of the normal state behavior below $T_c$.}
\label{fig:Hcond}
\end{figure}

It is instructive to look at the Hall conductivity $\sigma_{xy}$, normalized to $B$,  
as shown in Fig.\ \ref{fig:Hcond}, since $\rho_{yx}=R_H B$, $\sigma_{xy}/B$ is
independent of $B$ in the normal state above 90 K and roughly follows a
$\sigma_{xy} \propto T^{-3}$ temperature dependence. This trend extends into
the vortex-liquid region at $B=13$\ T, followed by a gradual increase of the
exponent as $T$ is reduced. At lower magnetic fields, however, a negative
contribution appears to gain importance over the extrapolated (broken line
in Fig.\ \ref{fig:Hcond}) normal state behavior, and at $B<3$~T leads to a sign
change of $\sigma_{xy}$. With the further $B$ decrease, a third, positive
contribution sets in sharply at low fields and leads to the second
sign reversal of $\sigma_{xy}$ (and $\rho_{yx}$) at $B \le 1$\ T.
Thus, the delicate interplay of these three contributions evokes the complex 
features and sign reversals of the Hall effect in YBCO. 

The Hall conductivity may be decomposed into
\begin{equation}
\sigma_{xy}=\sigma_{xy}^N+\sigma_{xy}^S+\sigma_{xy}^P,
\label{eq:Hall}
\end{equation}
where  $\sigma_{xy}^N$ represents a quasiparticle or vortex-core contribution,
associated with the normal-state excitations, $\sigma_{xy}^S$ a
superconducting contribution, resulting from  hydrodynamic vortex effects
and superconducting fluctuations \cite{VANO95,DORS92,TROY93,KOPN93,NISH97},
and $\sigma_{xy}^P$ allows for a possible pinning-dependence of $\sigma_{xy}$.
The sign of $\sigma_{xy}^N$ is the same as that of the normal-state Hall effect,
but the sign of $\sigma_{xy}^S$ depends on details of the Fermi surface
\cite{DORS92,TROY93,KOPN93,NAGA98}. The initially proposed pinning
independence of the Hall conductivity \cite{VINO93,LIU95} implies that
$\sigma_{xy}^P=0$. 

The results presented in Fig.\ \ref{fig:Hcond} can be understood according to
Eq.(\ref{eq:Hall}) as follows: $\sigma_{xy}^N \propto B$ dominates at the high $B$
range and follows the extrapolation from the normal state (broken line in
Fig.~\ref{fig:Hcond}). Its increase above the extrapolation below $T_c(B)$
indicates reduced quasiparticle scattering in the superconducting state
\cite{HARR94}. The contribution $\sigma_{xy}^S <0$ is roughly $\propto 1/B$
in fields of a few tesla, and is commonly associated with the hydrodynamic
flux-flow effects \cite{KOPN96}. In fields $B<1$\ T , however, the $1/B$
dependence is strongly violated and rather approaches a $B$-linear behavior.
Only the fluctuation models \cite{NISH97} allow for natural explanation of the
observed $B$ dependence of $\sigma_{xy}^S$, shown in Fig.\ \ref{fig:Hcond}
(excluding the second sign reversal)  \cite{PUIC00}.

\begin{figure}[tb]
\includegraphics[clip,width=6.5cm]{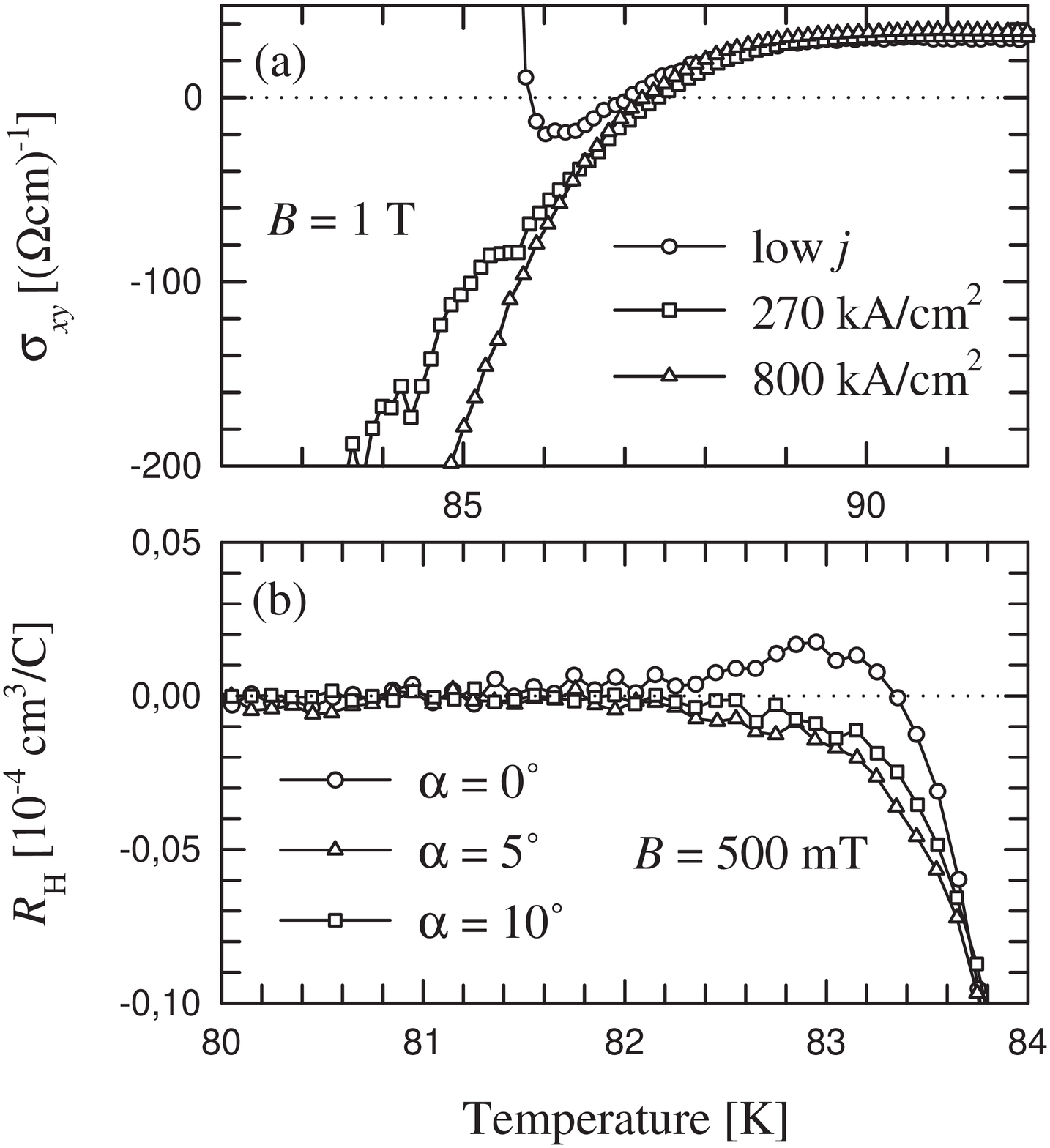}
\caption{Disappearance of the Hall effect's second sign reversal in YBCO films.
(a) Hall conductivity versus temperature for various current densities with
$B=1$~T, parallel to the film $c$-axis.
(b) Hall coefficient ($j=250$~A/cm$^2$) in oblique magnetic fields oriented
at the angle $\alpha$ relative to the film $c$-axis and to the twin boundaries.}
\label{fig:anom}
\end{figure}

The third (positive) contribution that is responsible for the second sign
reversal of $R_H$ (see Fig.~\ref{fig:Hall}), and that is evident from
Fig.\ \ref{fig:Hcond}, is attributed to $\sigma_{xy}^P$.
Kopnin {\it et al.} \cite{KOPN99} have proposed that $|\sigma_{xy}^P|$
can exceed $|\sigma_{xy}^S|$, leading eventually to an additional sign
reversal of the vortex Hall effect due to strong pinning. Ikeda \cite{IKED99}
has stressed that in the scenario of vortex-glass fluctuations, the sign of
 $\sigma_{xy}^P$ does depend on the dimensionality of the pinning, namely
${\rm sgn}(\sigma_{xy}^P)={\rm sgn}(\sigma_{xy}^S)$ for nearly
three-dimensional systems with disordered point-like pinning sites, and
${\rm sgn}(\sigma_{xy}^P) \ne {\rm sgn}(\sigma_{xy}^S)$ when line-like
pinning disorder dominates (Bose glass). The latter situation corresponds
to our YBCO films with $B$ oriented parallel to the twin-boundary planes.

\begin{figure}[tb]
\includegraphics[clip,width=6cm]{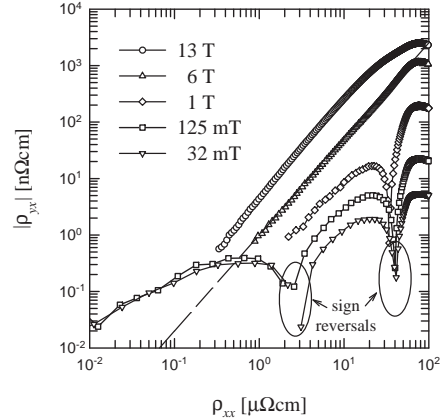}
\caption{Scaling of the transverse (Hall) and longitudinal resistivies in
various magnetic fields $B$ parallel to the film $c$-axis. The data for
1, 6, and 13 T have been truncated at low dissipation. The broken line
represents a fit to the 6 T data according to
$|\rho_{yx}| \propto \rho_{xx}^{1.7}$.}
\label{fig:scaling}
\end{figure}

Two additional experiments allow us to test the above notion on the origin
of $\sigma_{xy}^P$. The results are shown in Fig.~\ref{fig:anom}.
Figure~\ref{fig:anom}(a) presents $\sigma_{xy}$ dependence on temperature
for $B=1$~T and various transport current densities. In high current densities,
the pinning force is overcome by large Lorentz forces on the vortices 
and the sharp upturn of $\sigma_{xy}$, seen for low $j$ (see also
Fig.~\ref{fig:Hcond}), is canceled. Thus $\sigma_{xy}$ remains negative below
$T_c$ and rapidly decreases at lower temperatures. In this case,
$\sigma_{xy}$ strongly resembles the behavior seen in single crystals
\cite{DANN98}. Tilting  $B$ field off the $c$-axis at a small angle
$\alpha \le 10^\circ$, as shown in Fig.~\ref{fig:anom}(b), also leads to the
disappearance of the Hall effect's second sign reversal. In this case, pinning
is changed from line-like pinning along the twin boundary planes
($\alpha = 0^\circ$) to a reduced dimensionality pinning at an oblique field.
At $\alpha < 10^\circ$  we do not expect to reach point-like pinning, but
rather pinning along short segments, where the vortices run parallel to the
twin boundaries. Simultaneously, the longitudinal conductivity $\sigma_{xx}$
is not changed in oblique fields within the resolution of our measurement.
In this respect the impact of the $B$ tilt on $\sigma_{xy}$ is remarkable.
The results presented in Fig.~\ref{fig:anom} were obtained on films deposited
on different substrates, what confirms that the second sign reversal
is not associated with the film fabrication and/or sample inhomogeneities,
or due to a particular percolation path for the Hall current. 
It should be finally noted that our results are compatible with a study on
clean YBCO single crystals that reveal a significant difference in
$\sigma_{xy}$ measured at $B$ tilted at $\alpha=0^\circ$ and
$\alpha=4^\circ$, respectively, although the second sign reversal was there
not observed \cite{DANN98}.

The refutation of the pinning independence of $\sigma_{xy}$ concept by our
experiments immediately suggests a test of the scaling law -- the another
prediction that has been derived in the same theoretical context.
Previous work was limited to moderate magnetic fields and
Fig.\ \ref{fig:scaling} shows (broken line) the commonly observed scaling
$|\rho_{yx}| \propto  \rho_{xx}^{1.7}$ at, e.g., $B=6$\ T over more than four
orders of magnitude of $\rho_{yx}$. In fields $B \le 1$\ T, and at
temperatures, where  $\rho_{yx}<0$, scaling can be found only on a limited
range. Finally, in the strong pinning regime, where the second sign reversal
is observed and $\rho_{yx}>0$, the resistivities are related roughly as
$\rho_{yx} \propto  \rho_{xx}$, in sharp contrast to $\beta = 2$
\cite{VINO93,LIU95}, but not incompatible with the scaling law near a
vortex-glass transition \cite{DORS92a}.

Our experimental results impose several new, additional constraints on
models that attempt to explain the anomalous Hall effect in HTS.
Wang {\it et al.} \cite{WANG91,WANG94} associate the negative Hall anomaly
with the backflow current due to pinning. The increase of the negative
anomaly in low magnetic fields seems to support their model, but the second
sign change and the breakdown of the scaling in the limit of strong pinning
are hardly compatible. Several models based on the time-dependent
Ginzburg-Landau theory do not incorporate pinning effects and, thus, cannot
be applied to the low dissipation limit. Van Otterloo {\it et al.}
\cite{VANO95} and Kopnin \cite{KOPN96} have argued that the double sign
reversal is an intrinsic electronic phenomenon. While this seems to be
applicable for the highly anisotropic HTS, it is not the case for our YBCO
films. Ao \cite{AO98} has considered vortex lattice defects as the origin
of the sign reversal and predicted that the negative anomaly can disappear
in strong pinning and near a glass state. This latter notion is in agreement
with our experimental results, but contrary to theoretical predictions,
we did observe deviations from $\sigma_{xy} \propto 1/B$ dependence even
in the range where $\sigma_{xy}<0$. It has been recently pointed out by
Kopnin and Vinokur  \cite{KOPN99} that the initially proposed pinning
independence of $\sigma_{xy}$, and $|\rho_{yx}| \propto  \rho_{xx}^{2}$
is not valid in strong pinning case at twin boundaries, in accordance with
our observations. The dependence of $\sigma_{xy}^P$ on the dimensionality
of the vortex-pinning disorder, proposed by Ikeda \cite{IKED99}, is
supported by our results. Finally, D'Anna {\it et al.} \cite{DANN00} have
interpreted their scaling results on YBCO single crystals with
a percolation model that does predict $|\rho_{yx}| \propto  \rho_{xx}^{\beta}$
with $\beta=2$ for $B$ oriented parallel to the twin boundaries and
$\beta=1.4$ for slightly oblique fields. The latter exponent is similar to our
result when $\rho_{yx}<0$ in Fig.\ \ref{fig:scaling} but is incompatible with
our findings in the strong pinning limit.

In summary, we have measured the Hall effect of YBCO thin films in very
low magnetic fields and found a double sign reversal that is associated with
strong pinning along twin boundaries. The second sign reversal does vanish
at high current densities when the pinning is reduced, and also in slightly
oblique $B$ fields when the vortices transform from Bose glass to a vortex
glass. The scaling law proposed for the Hall and longitudinal resitivities
breaks down in the regime of the second sign reversal. Our data endorse
that the Hall conductivity is significantly influenced by vortex pinning.

We are grateful to R. Ikeda for very helpful correspondence and to Y. Matsuda
and N. Kopnin for stimulating discussions. This work was supported by the
Fonds zur F\"orderung der wissenschaftlichen Forschung, Austria, and by
the National Science Foundation grant DMR-0073366, Rochester.

\end{document}